\documentclass[a4paper]{jpconf}
\usepackage{graphicx}
\begin{document}
\title{Optical conductivity and superconductivity in LaSb$_2$}

\author{J F DiTusa$^1$, V Guritanu$^2$, S Guo$^1$, D P Young$^1$, P W
Adams$^1$, R G Goodrich$^1$, J Y Chan$^3$ and D van der Marel$^2$}

\address{$^1$ Department of Physics and Astronomy, Louisiana State
University, Baton Rouge, LA, 70803, USA} 
\address{$^2$ D\'{e}partement de Physique de la Mati\`{e}re Condens\'{e}e, 
Universit\'{e} Gen\`{e}ve, CH-1211 Gen\`{e}ve 4, Switzerland} 
\address{$^3$ Department of Chemistry, Louisiana State
University, Baton Rouge, LA 70803, USA}

\ead{ditusa@phys.lsu.edu}

\begin{abstract}
We have measured the resistivity, optical conductivity, and magnetic
susceptibility of LaSb$_2$ to search for clues as to the cause of the
extraordinarily large linear magnetoresistance and to explore the
properties of the superconducting state. We find no evidence in the
optical conductivity for the formation of a charge density wave state
above 20 K despite the highly layered crystal structure. In addition,
only small changes to the optical reflectivity with magnetic field are
observed indicating that the MR is due to scattering rate, not charge
density, variations with field. Although a superconducting ground
state was previously reported below a critical temperature of 0.4 K,
we observe, at ambient pressure, a fragile superconducting transition
with an onset at 2.5 K.  In crystalline samples, we find a high degree
of variability with a minority of samples displaying a full Meissner
fraction below 0.2 K and fluctuations apparent up to 2.5 K. The
application of pressure stabilizes the superconducting transition and
reduces the anisotropy of the superconducting phase.
\end{abstract}

\section{Introduction}
\begin{figure}[h]
\begin{minipage}{18pc}
\includegraphics[width=17pc,bb=100 50 630 500,clip]{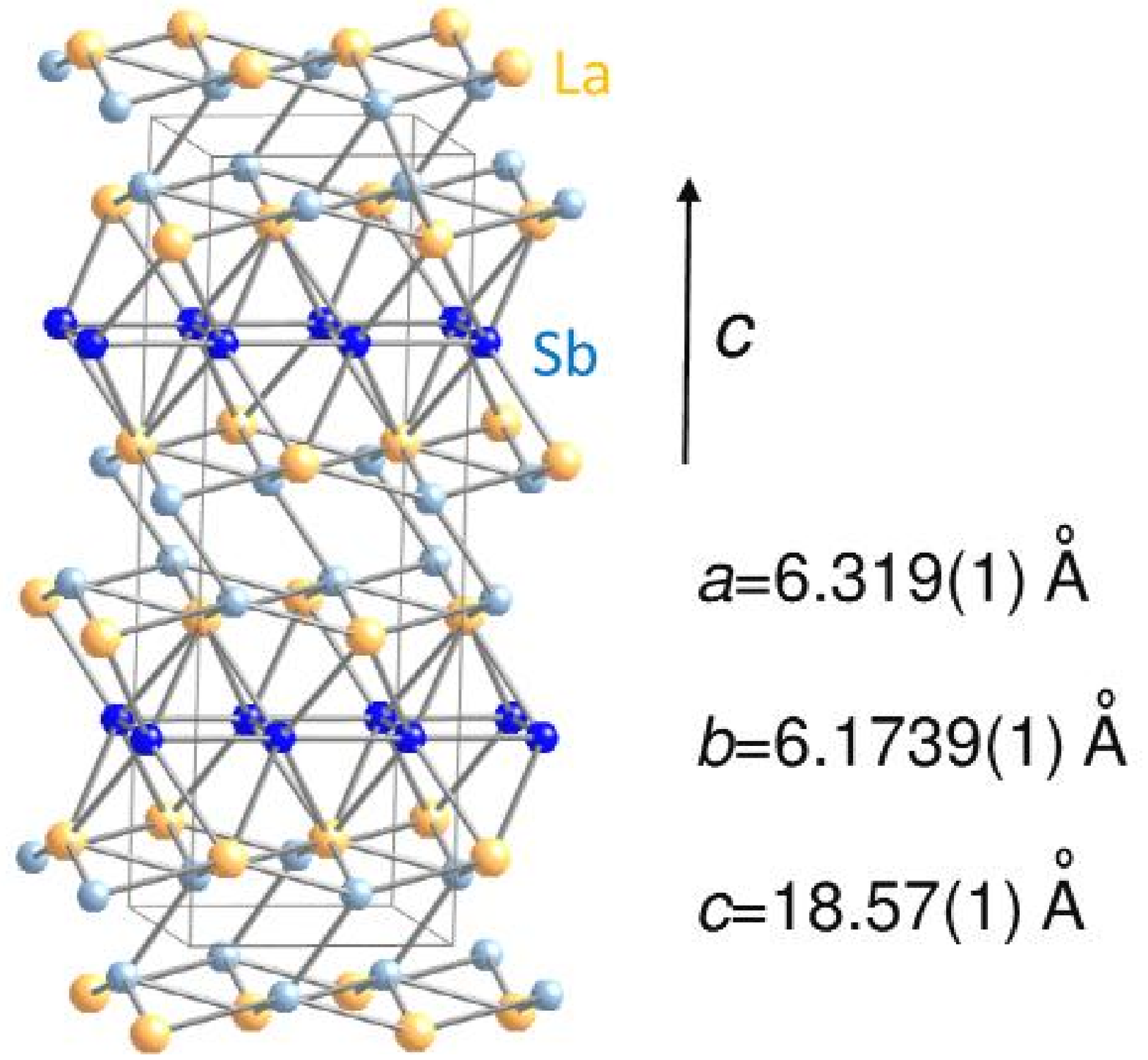}
\caption{\label{fig:struct}Lattice structure of LaSb$_2$. The layered
crystal structure highlighted by flat sheets of Sb (blue) separated by
the La (orange) and Sb interlayers. The unit cell is indicated by the
box. }
\end{minipage}\hspace{2pc}%
\begin{minipage}{18pc}
\includegraphics[width=12pc,angle=90,bb=50 145 485 740,clip]{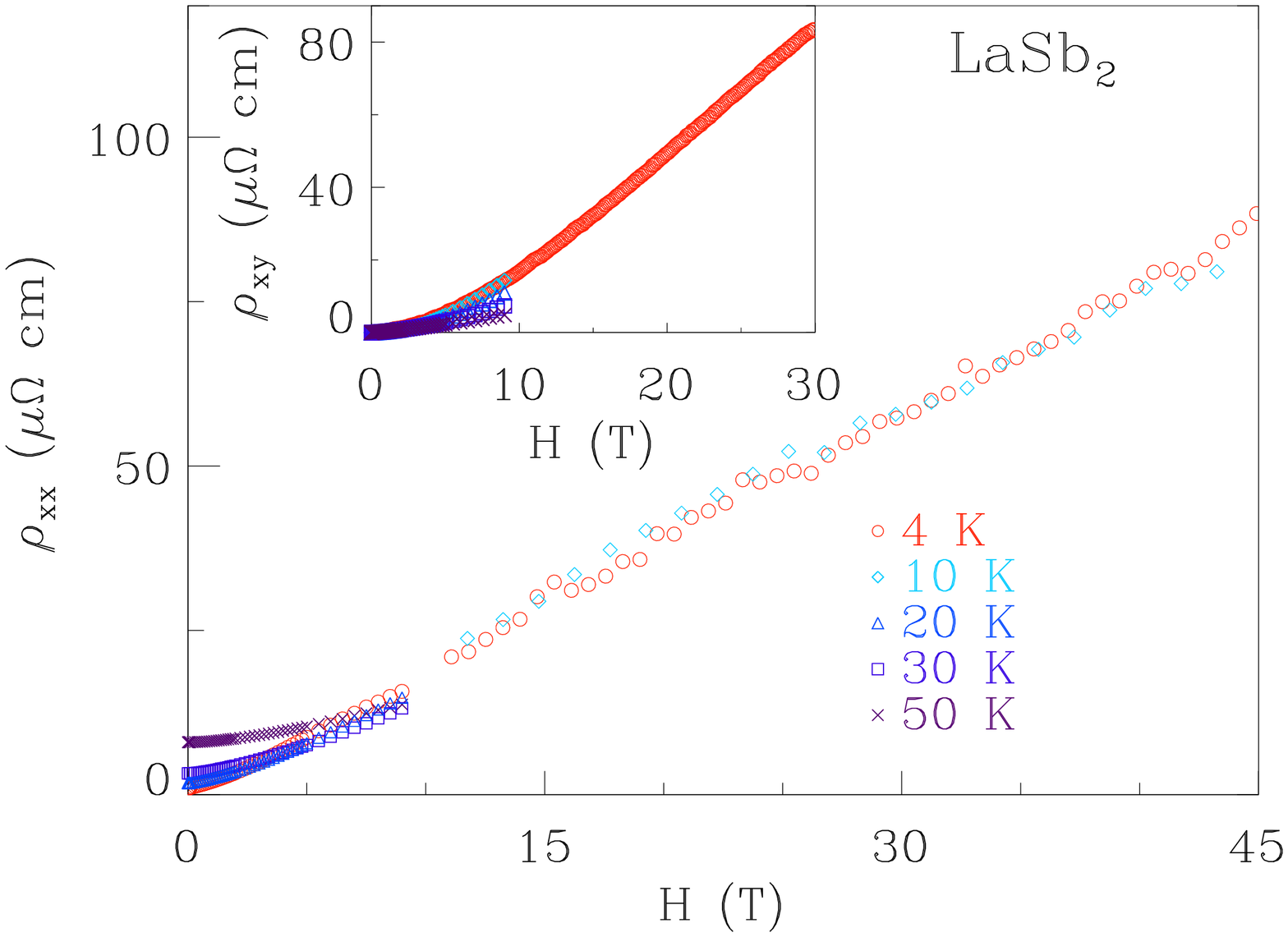}
\caption{\label{fig:mrhall}Magnetoresistance and Hall
  effect. Resistivity, $\rho_{xx}$, vs transverse magnetic field, $H$,
  for single crystal LaSb$_2$ at temperatures identified in
  the figure. Inset: Hall resistivity, $\rho_{xy}$ vs $H$ for the same
  temperatures (symbols same as in main frame).  }
\end{minipage}
\end{figure}

The observation of a large linear magnetoresistance, MR, in the
diantimonide, $R$Sb$_2$ ($R=$La-Nd, Sm), family of compounds has
stimulated recent interest because of the difficulty in modeling such
behavior.  These materials are layered compounds that crystallize in
the orthorhombic SmSb$_2$ crystal structure demonstrated in
Fig.~\ref{fig:struct}\cite{wang}. This structure is characterized by
La/Sb layers that alternate with flat, rectangular, sheets of Sb atoms
stacked along the crystallographic c-axis. All of the $R$Sb$_2$
compounds are metallic with LaSb$_2$ reported to be superconducting,
SC, with a critical temperature, $T_c$, of 0.4 K\cite{ott} that is not
yet fully characterized. For these reasons we investigated LaSb$_2$ to
explore the mechanism for the large MR and its SC properties.

The MR of LaSb$_2$ reproduced in Fig.~\ref{fig:mrhall} is linear in
field above 3 T\cite{budko}, is highly anisotropic, and does not
saturate in fields of up to 45 T. These features led to the suggestion
that LaSb$_2$ is useful as a magnetoresistive sensor for high field
applications\cite{younglasb2}.  Several mechanisms have been proposed
over the past 35 years to explain anomalously linear MRs which were
observed in disordered semiconductors\cite{husmann}, two-dimensional
(2D) heterostructures\cite{stormer}, and elemental
metals\cite{falikov,kapitza}. The purported causes include the
nucleation of charge density wave (CDW) distortions with MR due to
quantum fluctuations about the CDW\cite{castroneto} ground state
and/or a magnetic breakdown of the CDW gap\cite{wilson}, high field
quantization effects\cite{abrikosov}, or singular scattering
mechanisms\cite{young}.  While LaSb$_2$ does not appear to be in the
extreme quantum limit at the fields of Fig.~\ref{fig:mrhall}, and
there is no indication that conditions are favorable for strongly
singular scattering, it remains possible that there is a CDW ordering.
This possibility is supported by the fact that NbSe$_2$ and TaSe$_2$,
two prototypical CDW systems\cite{moncton}, have similar crystal
structures and linear MRs\cite{morris,naito}. However, measurements of
the resistivity, $\rho$, showed no transitions or unusual temperature,
$T$, dependencies that one usually associates with a CDW
transition\cite{budko} and photoemission\cite{acatrinei} and neutron
diffraction measurements\cite{kurtz} did not reveal any phase
transitions.

Here, we present optical ellipsometry, reflectivity, $\rho$, and
magnetic susceptibility, $\chi$, measurements of single crystalline
LaSb$_2$. In agreement with the previous measurements outlined above,
our optical measurements reveal no features indicative of CDW
formation. We find an extraordinarily wide transition to an
anisotropic SC phase that sharpens with pressure and which suggests
that SC phase fluctuations determine $T_c$.

\section{Experimental details}
Single crystals of LaSb$_2$ were grown from high purity La and Sb by
metallic flux methods that had typical dimensions 0.5 by 0.5 by 0.02
cm. The crystal structure displayed in Fig.~\ref{fig:struct} was
determined from single crystal x-ray diffraction. Optical experiments
were performed using spectroscopic ellipsometry at 0.75 - 3.7 eV in
combination with reflectivity measurements from 7 meV to 0.85
meV. Because there is no significant difference between the 300 and 20
K reflectivity spectra in the mid-infrared energy range, the
ellipsometry data were obtained at room temperature. All measurements
were performed on freshly polished crystals introduced into a vacuum
cryostat for the $T$-dependent measurements and kept in a flow of dry
nitrogen for room $T$ experiments in order to avoid the contamination
of the surface. To obtain the absolute value of the reflectivity a
reference gold layer was evaporated {\it in situ} on the sample
surface. From the frequency, $F$, dependent reflectivity and
ellipsometry data we derived the real part of the conductivity using a
Kramers-Kronig consistent variational fitting
procedure\cite{kuzmenko}. Reflectivity measurements were carried out
in $H$s of up to 7 T at a few selected $T$s. ac susceptibility
measurements were performed in a commercial SQUID magnetometer down to
1.75 K and in a dilution refrigerator ac $\chi$ probe for $T\ge 50$
mK. Hall effect and $\rho$ measurements were performed on single
crystals with electrical contacts made using silver epoxy employing
lock-in techniques at 17 or 27 Hz at $T$s between 0.05 and 300 K. Some
measurements reported here were made at the National High Magnetic
Field Laboratory in continuous fields of up to 45 T (35 T) for $\rho$
(Hall effect) measurements. The crystals studied here had residual
$\rho$ ratios of between 70 and 90 for $T$s between 300 and 4 K.

\section{Experimental Results}

\begin{figure}[h]
\begin{minipage}{18pc}
\includegraphics[width=13pc,angle=90,bb=50 140 485
740,clip]{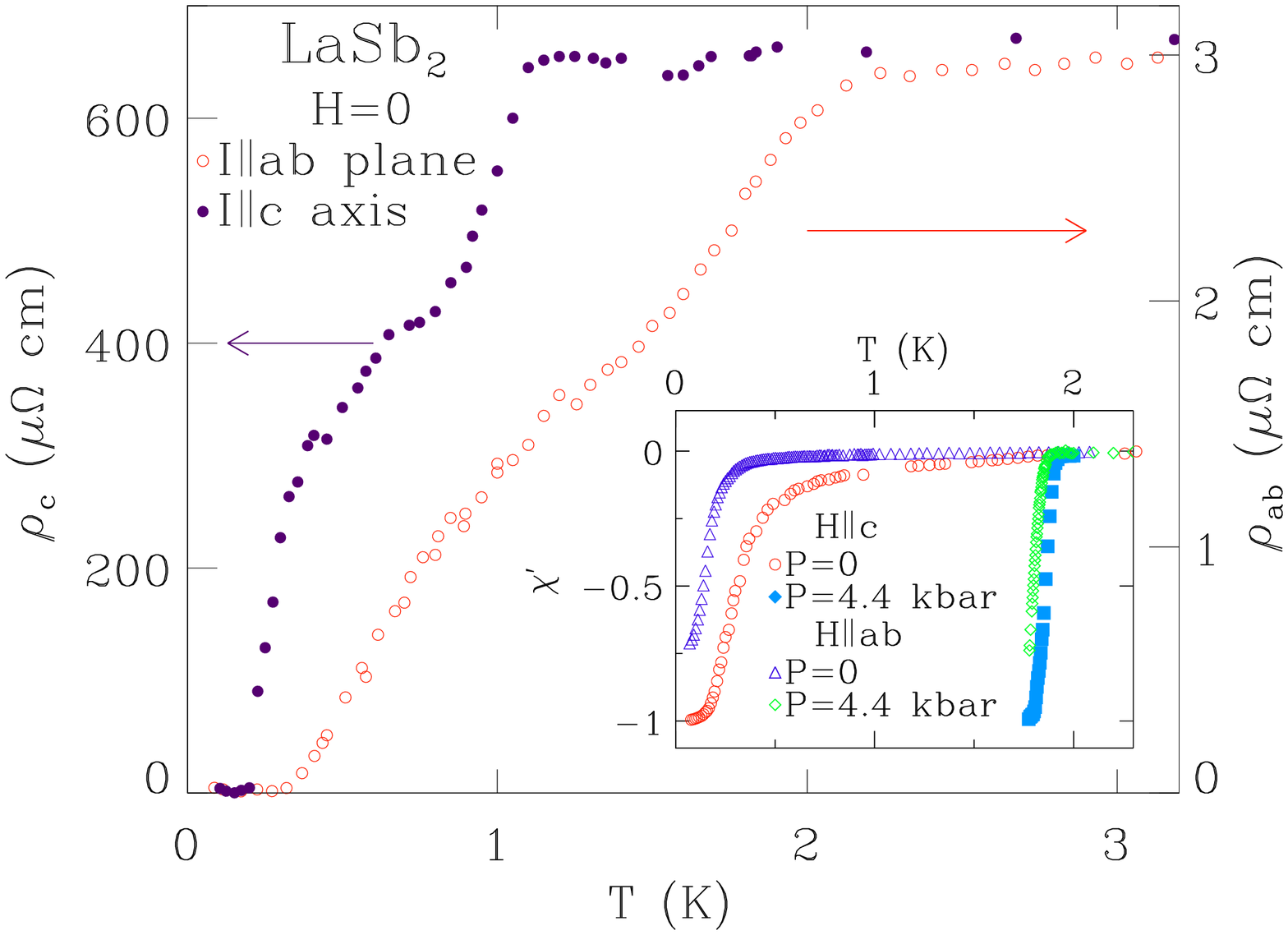}
\caption{\label{fig:acchi}Resistivity and Magnetic susceptibility. The
resistivity for currents parallel to the c-axis, $\rho_c$ and the ab
planes, $\rho_{ab}$ vs temperature, $T$, across the superconducting
transition. Inset: The real part of the ac magnetic susceptibility,
$\chi'$, vs $T$ at zero and 4.4 kbar applied hydrostatic pressure for
two ac excitation magnetic field orientations.  }
\end{minipage}
\hspace{2pc}%
\begin{minipage}{18pc}
\includegraphics[width=14pc,bb=0 0 205 246,clip]{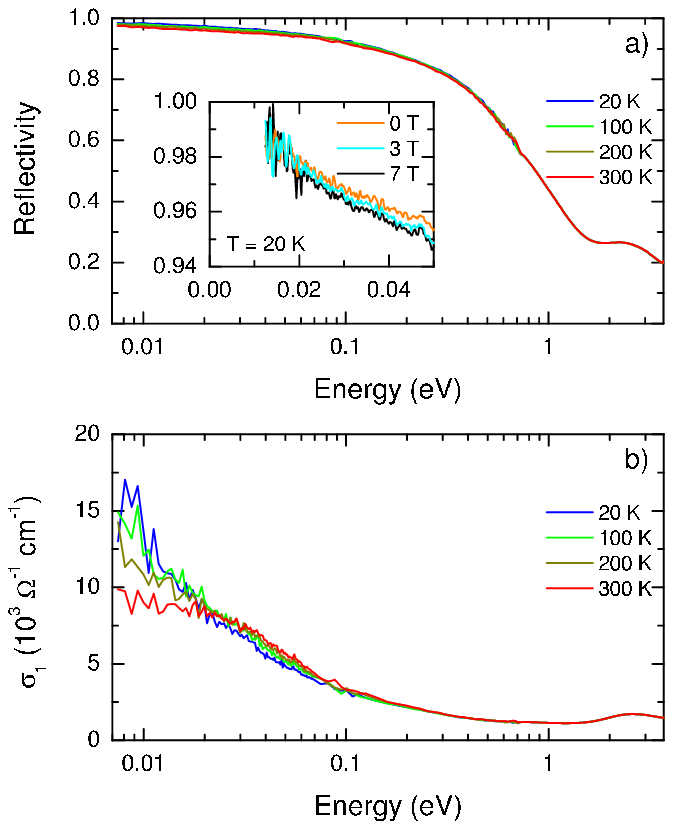}
\caption{\label{fig:optical}Optical response of LaSb$_2$. The
reflectivity (a) and optical conductivity, $\sigma_1$, (b) at
temperatures identified in the figure.  Inset: Reflectivity at 20 K in
magnetic fields of 0, 3 and 7 T. }
\end{minipage} 
\end{figure}

The transverse MR, $\rho_{xx}(H)$, and Hall resistivity, $\rho_{xy}$,
measured with currents parallel to the ab planes presented in
Fig.~\ref{fig:mrhall} are similar to previous
measurements\cite{budko,younglasb2} displaying a large, highly linear
MR and a large Hall constant. With the Hall carrier density, $n$,
estimated to be $2\times 10^{20}$ cm$^{-3}$ and using the
effective mass determined from previous de Haas-van Alphen
experiments\cite{lasb2dhva} of $m^* = 0.2$ times the bare electron
mass, we estimate the mean free path, $\ell$, to be 3.5 $\mu$m for
carriers moving along the ab planes. The resistivity along the c-axis,
$\rho_c$ is up to 200 times larger at 4 K as can be seen in
Fig.~\ref{fig:acchi}.

The optical reflectivity spectra of LaSb$_2$ at several $T$s is
presented in Fig.~\ref{fig:optical}. All spectra show a typical
metallic behavior with a high value of the reflectivity at low $F$ and
a well defined plasma edge at 1.7 eV. The conductivity, $\sigma_1$,
spectra obtained from the reflectivity and ellipsometry measurements
are displayed in Fig.~\ref{fig:optical}b. These data show a small $T$
dependence and no interband transition appears to be present at low
energy. At higher energy a peak centered at about 2.5 eV is attributed
to an interband transition. At low $F$, $\sigma_1$ is
featureless and displays no features which can be associated with the
opening of a CDW gap. The normalized reflectivity versus photon energy
for three applied magnetic fields at $T = 20$ K is shown in the inset
of Fig.~\ref{fig:optical}a. We see no systematic trends in these data
outside of the measurement error as $H$ is increased. The data agree
with simple estimates of the field induced changes to the reflectivity
based upon the MR of Fig.~\ref{fig:mrhall} and the Hagen-Rubens
formula. Our data rule out any significant change to the electronic
structure of LaSb$_2$ with magnetic field at energies above 10 meV.

The wide transition into the SC phase as probed by $\rho$ and ac
$\chi$ measurements is displayed in Fig.~\ref{fig:acchi}. Both of
these quantities show a very wide, anisotropic, phase transition where
$\rho$ decreases and the real part of the ac magnetic susceptibility,
$\chi'$, inset, displays diamagnetism below 2.5 K.  Here $\chi'$ is
shown for a crystal displaying a complete Meissner fraction,
$\chi'=-1$, for ac excitation fields, $H_{ac}$, along the c-axis for
$T < 0.25$ K. The application of moderate pressure, represented in the
inset to Fig.~\ref{fig:acchi} by data at 4.4 kbar, significantly
sharpens the transition so that by 1.7 K $\chi'=-1$ for
$H_{ac}\parallel c$.

\section{Discussion and Conclusions}
Our measurements of $\sigma_1$ of LaSb$_2$ effectively rule out the
formation of a CDW phase above 20 K showing no unusual changes from a
typically metallic form. In addition, we see no significant changes in
the reflectivity with magnetic field.  Thus, our data suggest that the
MR is a scattering induced effect which is not related to the
formation of a CDW induced energy gap or variations of the electronic
structure with $H$. In addition, our measurements reveal an unusual,
anisotropic, SC transition that is very broad in $T$ that is
incomplete down to our lowest $T$s in the majority of crystals. The
application of $P$ sharpens the transition dramatically, inducing a
full SC transition. Similar SC phase transitions have been previously
observed in layered materials such as 2H-TaS$_2$ where the width of
the transition was associated with the competition of a CDW
phase\cite{nagata}. In LaSb$_2$ we have found no indication of a CDW,
or other energy gap inducing, phases that compete with
superconductivity either in $\sigma_1$ above 20 K nor in $\rho(T)$ at
any $T$. This suggests a different cause for the wide SC transition we
observe. Because the transition sharpens with pressure, and because of
the long $\ell$ for charge carriers moving parallel to the ab planes,
we do not believe impurity or crystal defects to be important. Instead
we suggest that the SC phase in LaSb$_2$ at $P=0$ may be phase
fluctuation limited as in the underdoped high $T_c$ SC cuprates.

\subsection{Acknowledgments}
JFD, DPY, and JYC acknowledge support from the NSF through DMR 084376,
0449022, and 0756281. PWA acknowledges support from the U.S. DOE
through DE-FG02-07ER46420. Work at the NHMFL was performed under the
auspices of the NSF and the state of Florida.

\end{document}